# Investigation of stimulated dynamics of vortex-matter in high-temperature superconductors

J.G. Chigvinadze*, J.V. Acrivos**, S.M. Ashimov*, A.A. Iashvili*, T. V. Machaidze*, G.I. Mamniashvili*, Th. Wolf***

*E. Andronikashvili Institute of Physics, 0177 Tbilisi, Georgia
**San Jose' State University, San Jose' CA 95192-0101, USA
*** Forschungszentrum Karlsruhe, Institut für Festkörperphysik, 76021 Karlsruhe, Germany

*A simple mechanical method for the investigation of Abrikosov vortex lattice stimulated dynamics in superconductors has been used. By this method we studied the action of pulsed magnetic fields on the vortex lattice and established the resulting change of the course of relaxation processes in the vortex matter in high-temperature superconductors. This method can be used for investigation of phase transitions in vortex matter both high-temperature and exotic superconductors.*

## INTRODUCTION.

Investigation of vortex state stimulated dynamics in high-temperature superconductors (HTSC) is one of the most important problems from the point of view of practical applications in the new generation of electronic devices operating on the basis of HTSC materials, as well as, from the point of view of fundamental problems connected with high-temperature superconductivity [1]. It is clear that vortex states in superconductors are complicated and multiform, but they are exceptionally important for understanding the behavior of high-temperature superconductors in electromagnetic fields and under the current loads [2]. The critical temperature $T_c$ of the superconducting transition of HTSC cuprates is so high that they remain superconductive at temperatures, at which the thermal fluctuations play a noticeable role, as their energy become comparable with the elastic energy of vortices and also with the pinning energy [3]. It creates prerequisites for phase transitions [4-18]. In the type-II HTSC superconductors the number of vortices, "particles" in the vortex matter may be changed in a very wide range, several orders of magnitude by varying the magnetic field, **B**. As the interaction between the vortices is changed, thermal fluctuations in the HTSC vortex matter can be observed in a very wide temperature interval, on the ***B-T*** phase diagram. This gives a good possibility to study the disordered media – one of the central problems of condensed state physics. The investigation of vortex matter dynamics is of particularly topical interest in the case of HTSCs, which, because of their high critical temperatures and layered structure are characterized by a much higher mobility of the vortex lattice, as compared with the ordinary type-II superconductors, which is hampered by pinning connected with various defects of the crystal structure. These defects are the pinning centers that prevent the flow of magnetic flux and the energy dissipation, accompanying it. In the strongly anisotropic high-temperature superconductors, as it was shown in recent works [16-18], the critical current at the transition of *3D* vortices in the quasi-two-dimensional *2D* vortices can be sharply increased, making such anisotropic high-temperature superconductors, e.g. BiPbSrCaCuO system, useful for technical applications. Furthermore, the upper critical field $H_{c2}$, at which the superconductivity in these materials is destroyed, can reach 150 T [1]. This field is much higher, than in traditional type-II superconductors, used presently. This last circumstance makes the high-temperature superconductors promising materials for further technical developments.

In this work, the decisive role played by the Abrikosov vortex lattice stimulated dynamics is investigated by imposing a weak alternating or pulsed magnetic field on the external permanent magnetic field applied to the HTSC. The former causes the motion of the vortex continuum created by the latter. This, in turn, leads to a change of relaxation processes taking place in the vortex matter [8,9]. The investigation of these problems is of current concern for understanding the mechanisms of energy dissipation, which arises at the motion of vortex matter inside a superconductor, as well as for determining the limits of application of technical devices made on the basis HTSC.

## EXPERIMENTAL.

Among the methods of vortex matter stimulated dynamics investigation in superconductors one should single out two categories: The first are macroscopic methods of investigation, that include contactless mechanical measurement of pinning and dissipative processes in the superconductors [19], which, thanks to their high sensitivity, allow to detect the processes of detachment and motion of the most weakly fixed vortices at the very early stage of their motion, after detachment from the pinning centers. The second are microscopic methods, including the method of dephasing of the nuclear spin echo, that gives the spatial distribution of magnetic field in vortex cores [20], which is achieved by the action of magnetic pulses on the spin system of nuclei, located in normal cores of the Abrikosov vortex lattice. The mechanical method of investigation of the stimulated dynamics of the Abrikosov vortex lattice, proposed by us, is the development of a currentless mechanical method of the investigation of pinning [21,22] based on the measurement of the mechanical moment acting on the superconducting sample with axial symmetry, which is located in an external (transverse) magnetic field. Countermoments of pinning forces and of viscous friction, acting on the sample from the side of quantized vortex lines (Abrikosov vortices) were determined as in [23,24]. The isotropic ceramic $HoBa_2Cu_3O_{7-\delta}$ sample was prepared by the standard solid state reaction. Samples were made cylindrical with length $L$=13 and diameter $d$=6mm. Critical temperatures is $T_c$=92 K. Used samples were isotropic. This was checked by measurements of mechanical moments dependence on angle $\theta$ in the plane perpendicular to the rotation axis. Pulsed magnetic field with amplitude up to H≈400 Oe and duration about 30 µs was created by Helmholtz coils. The $HoBa_2Cu_3O_{7-\delta}$ sample was placed in the center of these coils. Pulsed magnetic field was parallel to the main outer stationary magnetic field used to create vortices. For the formation of such pulses was used standard generator and homebuilt power amplifier. The sensitivity of the method, obtained by the current voltage characteristics according to [25], is equivalent to $10^{-8}$ B·cm$^{-1}$. The angle of rotation of the sample, $\varphi_2$ is measured as a function of the angle of rotation of a torsion head, $\varphi_1$ transmitting the rotation to the sample by means of suspension, having a torsion stiffness $k$ ~4.10$^{-1}$ [dyn·cm] in the schematic diagram giving the geometry of the apparatus (FIG.1). The latter can be replaced, if necessary, by a less stiff or a stiffer one. The measurements were carried out at a constant speed of rotation of the torsion head, making $\omega_1$ = 1.8x10$^{-2}$rad/sec. Angles of rotation $\varphi_2$ and $\varphi_1$ were determined with an accuracy of ±4.6x10$^{-3}$ and ±2.3x10$^{-3}$ rad, respectively. The uniformity of the magnetic field strength along the sample were below $\Delta H/H$=10$^{-3}$. To avoid effects, connected with the frozen magnetic fluxes, the lower part of the cryostat with the sample was put into a special cylindrical Permalloy screen, reducing the earth magnetic field by a factor of 1200. After the sample, was cooled by liquid nitrogen to the superconducting state, the screen was removed, a magnetic field of necessary intensity $H$ was applied and the $\varphi_2(\varphi_1)$ dependence was measured. For carrying out measurements at different values of $H$, the sample was transferred to the normal state by heating it to $T$>$T_c$ at $H$=0 and only after returning sample and torsion head to the initial state $\varphi_2 = \varphi_1$, the experiment was repeated.

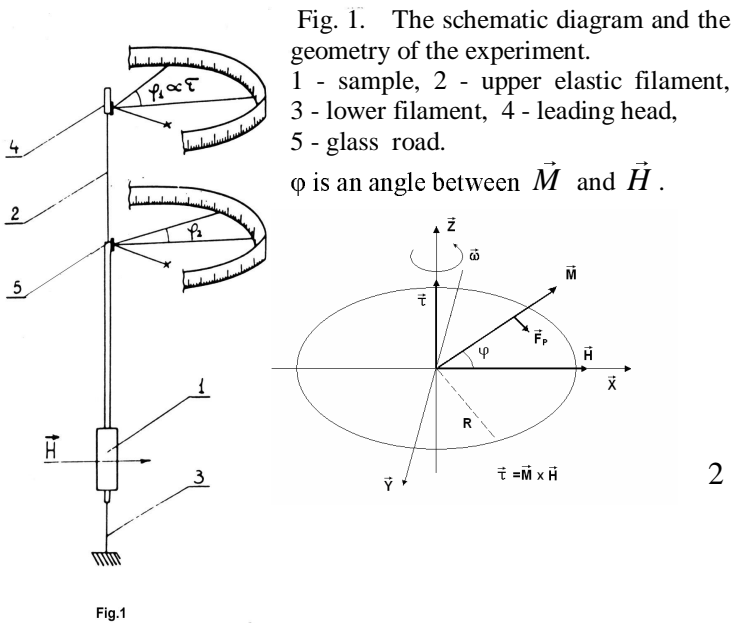

Fig. 1. The schematic diagram and the geometry of the experiment.
1 - sample, 2 - upper elastic filament, 3 - lower filament, 4 - leading head, 5 - glass road.
$\varphi$ is an angle between $\vec{M}$ and $\vec{H}$.

## RESULTS AND DISCUSSION.

The $\varphi_2$ versus $\varphi_1$ dependence, at T = 77 K, at various magnetic fields for $HoBa_2Cu_3O_{7-\delta}$ sample (length of a cylindrical sample L=13mm and the diameter d=6mm) is shown in FIG. 2. During rotation of the sample both in normal and superconducting states in the absence of external magnetic field ($H$=0) $\varphi_2$ versus $\varphi_1$ is linear and the condition $\varphi_1 = \varphi_2 = \omega t$ is satisfied. The character of the dependence $\varphi_2(\varphi_1)$ is changed significantly, when the sample is in magnetic fields $H$>$H_{c1}$ at T<$T_c$. Three distinct regions are observed (FIG.2, curve $H$=2900Oe). In the first (initial) region, the sample does not respond to the increase in $\varphi_1$, i.e., to the applied torque, $\tau$ (FIG. 1) increasing as $\tau = k(\varphi_1 - \varphi_2)$ or responds weakly.

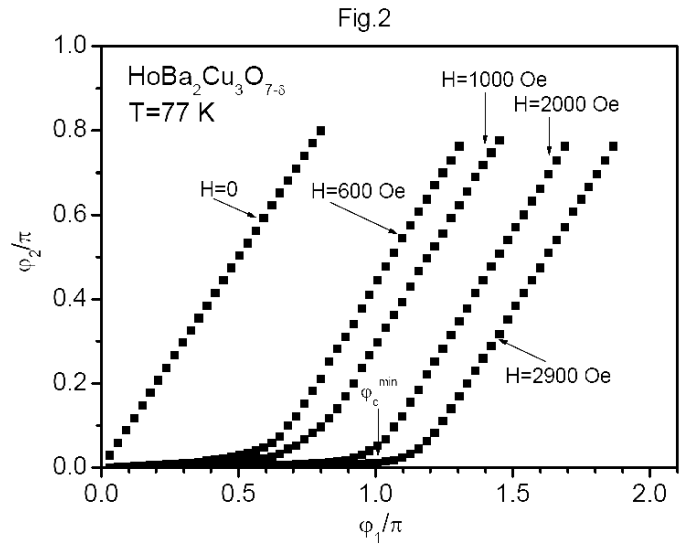

**FIG. 2:** *Dependence of the angle of rotation of the sample $\varphi_2$ on the angle of rotation of the leading head $\varphi_1$ in magnetic fields H= 0; 600; 1000; 2000 and 2900 Oe at T=77 K.*

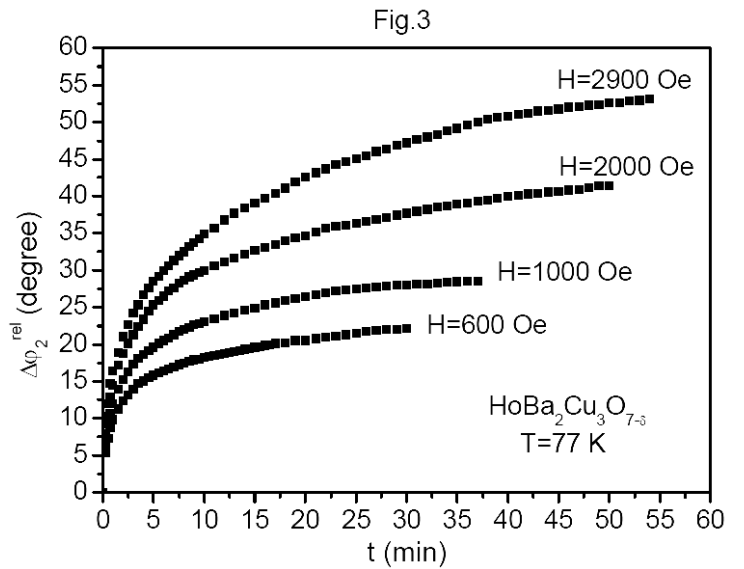

Fig.1

**FIG. 3:** *Time dependence of the angle of relaxation rotation of the sample at T=77 K in magnetic fields H= 600; 1000; 2000 and 2900 Oe.*

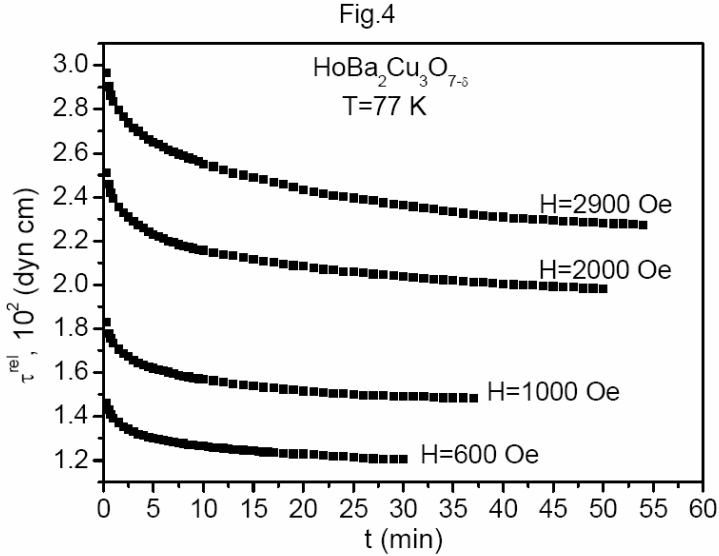

**FIG. 4:** *The dependence of the moment $\tau^{rel}$, connected with the relaxation rotation of the sample, at time t, at T=77 K in magnetic fields H=600; 1000; 2000 and 2900 Oe.*

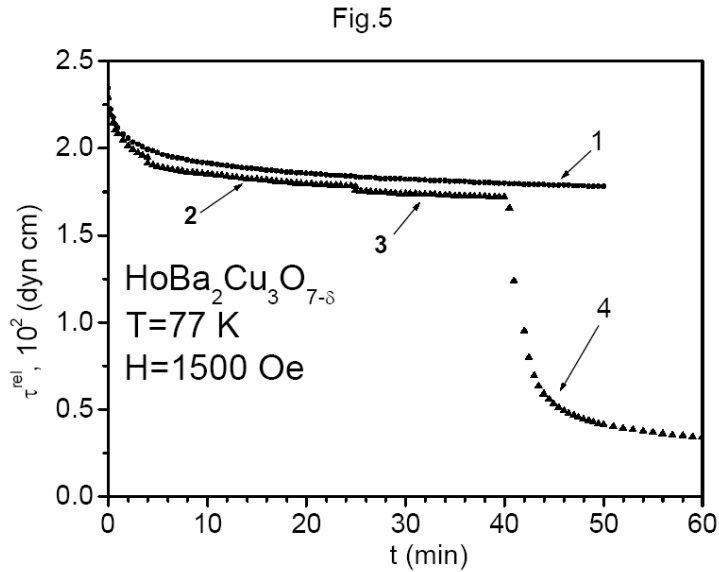

**FIG. 5:** *Dependence of the moment $\tau^{rel}$, connected with the relaxation rotation of the sample, at time t, at T = 77 K in a magnetic field H=1500 Oe. Curve 1 corresponds to the relaxation process without the action of magnetic pulse. Sections 2 and 3 - after the action of single magnetic pulse. Section 4 - continuous action of magnetic pulses, stimulating the acceleration of the relaxation process.*

Such behavior of the center can be explained by the fact that Abrikosov vortices are not separated from the pinning centers at small values of $\varphi_1 \propto \tau$ but if the sample is still turned slightly, this can be caused by elastic deformation of magnetic force lines beyond it or, possibly, by separation of the most weakly fixed vortices.

As soon as a certain critical value $\varphi_c^{min}$ dependent on $H$, the first region, undergoes a transition to the second region (FIG. 2), in which the velocity of the sample increases gradually with the increase of $\varphi_1$ resulting from the progressive process of separation of vortices from their corresponding pinning centers. One should expect that just in this region, in the rotating sample «a vortex fan» begins to unfold out, in which the vortices are distributed according to the instantaneous angles of orientation with respect to the fixed external magnetic field. In this case the values of the angles of orientation of separate vortex filaments are limited to $\varphi_{fr}$ and $\varphi_{fr} + \varphi_{pin}$, where $\varphi_{fr}$ is the angle, about which the vortex filament can be turned with respect to $H$ by the forces of viscous friction on the matrix of superconductor, and $\varphi_{pin}$ is the angle, about which the vortex filament can be turned with most strong pinning center, studied for the first time in [26]. The gradual transition (at high values of $\varphi_1$) to the third region, where the linear dependence $\varphi_2(\varphi_1)$ is observed, allows to determine the countermoments of pinning forces $\tau_p$ and $\tau_{fr}$, independently. Just in this region, when $\omega = \omega_1 = \omega_2$, the torque $\tau$, applied to uniformly rotating sample, is balanced by countermoments $\tau_p$ and $\tau_{fr}$. In particular, in the case of continuously rotating sample with frequency $\omega = \omega_1 = \omega_2$ one could find similarly to [27,28] an expression for the total braking torque $\tau$.

Indeed, if we consider in this case a vortex element $\vec{ds}$ moving with velocity $\vec{\upsilon}_\perp$ perpendicular to $\vec{ds}$, then the average force acting on this elements is

$$d\vec{f}_\upsilon = \vec{\upsilon}_\perp \eta ds + \frac{\vec{\upsilon}_\perp}{|\upsilon_\perp|} F_l ds$$

and the associated braking torque, exerted on the rotating specimen becomes:

$$d\vec{\tau} = \vec{r} \times d\vec{f}_\upsilon$$

where $\vec{r}$ is the vector pointing from the rotational axis to the vortex elements, $F_l$ is the pinning force per flux thread per unit length, and $\eta$ is the viscosity coefficient. For a cylindrical specimen of radius $R$ and height $L$ integrating over the individual contribution of all vortex gives a total braking torque $\tau$

$$\tau = \tau_p + \tau_0 \omega \qquad (1)$$

with

$$\tau_p = \frac{4}{3}\frac{BF_l}{\Phi_0}LR^3 ,$$

and

$$\tau_0 = \frac{\pi}{4}\frac{B}{\Phi_0}\eta LR^4 ,$$

where $B$ is the inductivity averaged over the sample, $\Phi_0$ is the flux quantum, $L$ is the height and $R$ is the radius of the sample.

If, in this region the torsion head is stopped, at the expense of relaxation processes, connected with the presence of the viscous forces acting on the vortex filaments, the sample will continue to rotate to the same direction (with decreasing velocity) until it reaches a certain equilibrium position, depending on the value of $H$. FIG. 3 shows the curves of time dependence of $\Delta\varphi_2^{rel}$ at the stopped leading head for HoBa$_2$Cu$_3$O$_{7-\delta}$ sample at T=77 K when H=600, H=1000, H=2000 and H=2900, respectively. According to the angle of relaxation rotation $\Delta\varphi_2^{rel}$, one can easily determine the countermoment of $\tau_{fr}$ forces, acting on vortex lines in the process of relaxation $\tau_{fr} = k\Delta\varphi_2^{rel}$. The dependence of $\tau^{rel}$ moment, connected with the relaxation rotation, on time t at T = 77 K in various magnetic fields is shown in FIG. 4. According to (1) the torque $\tau$ applied to the sample after its relaxation rotation and stopping ($\omega = 0$), will be balanced by the countermoment of only static pinning forces.

$$\tau^{st} = k[\varphi_1 - \varphi_2 - \Delta\varphi_2], \qquad (2)$$

where $\varphi_1$ and $\varphi_2$ are the values of angles at the moment of stopping the leading head. As it is seen from (1), the measurements of $\tau_p(H)$ dependence allows to determine the volume pinning force as well:

$$F_p = 3\tau_p / 4R^3 L \qquad (3)$$

If in the process of the sample relaxation rotation, we apply a pulsed magnetic field parallel to the applied external permanent magnetic field, the additional vortices, produced as a result of the magnetic pulse, will cause a separation of main vortices, produced by the permanent magnetic field, from the pinning centers. The latter vortices, in their turn, will cause an additional change of relaxation processes and thus, an abrupt decrease of the moment, connected with viscous forces $\tau_{fr}$. Curve 1 (FIG. 5) corresponds to the relaxation process of the rotation of a sample without action of a magnetic pulse. Curve 2 is registered after the action of one pulse (h~400 Oe, duration of the pulse is 30 μs), the action of which is accompanied by an abrupt decrease of $\tau^{rel}$ moment and by the further gradual decrease of $\tau^{rel}$ until the second pulse of the same value and duration is applied. In the case of the action of the second pulse, an abrupt decrease of the $\tau^{rel}$ moment and its further gradual decrease are observed. And finally, curve 4 corresponds to the continuous action of above-mentioned pulses with the repetition frequency, equal to 2.5 Hz. The time dependence of the relaxation moment $\tau^{rel}$ for curve 1 in the absence of the action of magnetic pulses and of the sections 2 and 3, is of logarithmic character, when using single pulses. When pulses are applied with a repetition frequency of 2.5 Hz, the relaxation law is changed and has an exponential character.

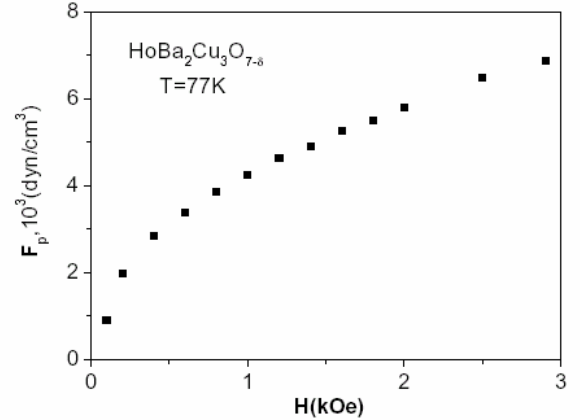

**FIG. 6:** *Dependence of the volume force of pinning $\vec{F}_p$ on the external magnetic field $H$.*

Using the $\varphi_2(\varphi_1)$ dependence and derived results (FIG. 2, 3, 4), the value of $\tau_p$ can be determined. Furthermore, using (3), the dependence of the volume force of pinning $F_p$ on the external magnetic field $H$ can be plotted. This dependence shows that with increase of external magnetic field $H$ the volume force of pinning $F_p$ increases (FIG. 6). By the investigation of the temperature dependence of the Abrikosov vortex lattice stimulated dynamics, using the method proposed by us, one can study the phase transitions in vortex matter, including the melting of Abrikosov vortex lattice and thus, determine the change of relaxation law, when studying the phase transitions in high temperature superconductors [29].

**CONCLUSION.**

A simple mechanical method for the investigation of Abrikosov vortex lattice stimulated dynamics in superconductors is proposed. By this method the action of pulsed magnetic fields on vortex lattice in HoBa$_2$Cu$_3$O$_{7-\delta}$ high-temperature superconductors is studied and the resulting change of the course of relaxation processes in the vortex matter of high-temperature superconductors has been established.

This method can be used for investigation of phase transitions in vortex matter both high-temperature and other superconductors. It is shown also that with the increase of outer magnetic field the pining force $F_p$ increases in field up to 3 KOe.

**ACKNOWLEDGEMENT.**
The work is supported by the grants of International Science and Technology Center (ISTC) G-389 and G-593.